\documentclass[english,showpacs,twocolumn,superscriptaddress,nofootinbib,aps]{revtex4-1}
\usepackage[T1]{fontenc}
\usepackage[latin9]{inputenc}
\usepackage{color}
\usepackage{babel}
\usepackage{float}
\usepackage{booktabs}
\usepackage{textcomp}
\usepackage{amsmath}
\usepackage{amssymb}
\usepackage{stmaryrd}
\usepackage{graphicx}
\usepackage{multirow}
\usepackage[unicode=true,pdfusetitle,
 bookmarks=true,bookmarksnumbered=false,bookmarksopen=false,
 breaklinks=false,pdfborder={0 0 1},backref=false,colorlinks=true]
 {hyperref}
\usepackage{ulem}



\begin{document}
\title{Production of Strange and Charm Hadrons in Pb+Pb Collisions at $\sqrt{s_{NN}}=$ 5.02 TeV}
\author{Wen-bin Chang}
\affiliation{School of Physics and Physical Engineering, Qufu Normal University, Shandong 273165, China}
\affiliation{Institute of Particle Physics and Key Laboratory of Quark and Lepton Physics (MOS), Central China Normal University, Wuhan 430079, China}
\author{Rui-qin Wang }
\affiliation{School of Physics and Physical Engineering, Qufu Normal University, Shandong 273165, China}
\author{Jun Song }
\affiliation{School of Physical Science and Intelligent Engineering, Jining University, Shandong 273155, China}
\author{Feng-lan Shao}
\email{shaofl@mail.sdu.edu.cn}
\affiliation{School of Physics and Physical Engineering, Qufu Normal University, Shandong 273165, China}

\author{Qun Wang}
\email{qunwang@ustc.edu.cn}
\affiliation{Department of Modern Physics, University of Science and Technology of China, Hefei, Anhui 230026, China}

\author{Zuo-tang Liang}
\email{liang@sdu.edu.cn }
\affiliation{Key Laboratory of Particle Physics and Particle Irradiation (MOE), Institute of Frontier and Interdisciplinary Science, Shandong University, Qingdao, Shandong 266237, China}

\begin{abstract}
    Using a quark combination model with the equal-velocity combination approximation, we study the production of hadrons with strangeness and charm flavor quantum numbers in Pb+Pb collisions at $\sqrt{s_{NN}}=$5.02 TeV. We present analytical expressions and numerical results for these hadrons' transverse momentum spectra and yield ratios. Our numerical results agree well with the experimental data available. The features of strange and charm hadron production in the quark--gluon plasma at the early stage of heavy ion collisions are also discussed.  
\end{abstract}
\maketitle

\section{Introduction\label{sec:Intro} }

It is well-known that the hadronic matter is expected to undergo a transition to the quark--gluon plasma (QGP), a~strongly coupled state of matter, at~high temperatures or baryon densities~\cite{Shuryak:1980tp,Braun-Munzinger:2008szb,Fukushima:2010bq,Borsanyi:2010bp,Bazavov:2011nk,Ding:2015ona}.
The search for the QGP and the study of its properties have long been the goals of high-energy heavy ion collisions~\cite{STAR:2005gfr,PHENIX:2004vcz,PHOBOS:2004zne,BRAHMS:2004adc}.
In heavy ion collisions, strange and heavy-flavor quarks are newly produced (or excited from the vacuum) and most of them are present in the whole stage of the QGP evolution. They interact strongly with the constituents of the QGP medium or they are a part of the QGP. Therefore, strange and heavy-flavor hadrons are usually regarded as special probes to the hadronization mechanism and properties of the QGP~\cite{Rafelski:1982pu,Shor:1984ui,vanHecke:1998yu,Matsui:1986dk,vanHees:2005wb,Mocsy:2007jz,He:2012df,Braun-Munzinger:2015hba,Batsouli:2002qf,Xu:2002hq,Lin:2003jy,Moore:2004tg}.


The relativistic heavy ion collider (RHIC) and the large hadron collider (LHC) have accumulated abundant experimental data on strange and charm hadrons~\cite{STAR:2008bgi,STAR:2015vvs,ALICE:2013xmt,PHENIX:2004ggw,STAR:2004ocv,ALICE:2012inj,ALICE:2012ab,STAR:2018zdy,ALICE:2018lyv,Radhakrishnan:2019gbl,Zhou:2017ikn,ALICE:2019fhe,ALICE:2018hbc,ALICE:2017thy}.
These data show a number of features in the production of strangeness~\cite{Adams:2006ke,STAR:2017sal,Zhou:2017ikn,Adam:2019koz,ALICE:2019fhe,ALICE:2021ptz} and baryons~\cite{STAR:2006uve,PHENIX:2006dpn,Blume:2011sb,ALICE:2014juv,Radhakrishnan:2019gbl,ALICE:2018hbc}. Many efforts have been made to understand hadron production mechanisms in theory and phenomenology~\cite{Kolb:2001qz,Shen:2010uy,Andronic:2007zu,Kuznetsova:2006bh,Andronic:2015wma,Greco:2003xt,Fries:2003vb,Hwa:2002tu,Shao:2004cn,Greco:2003vf,Oh:2009zj,Liu:2012tn,Lee:2007wr,He:2011qa,Plumari:2017ntm,Zhao:2018jlw,Cho:2019lxb,Wang:2013pnn}.
Hydrodynamic and thermal models~\cite{Kolb:2001qz,Shen:2010uy,Andronic:2007zu,Kuznetsova:2006bh,Andronic:2015wma} are commonly used to describe the production of strange hadrons.
For the production of heavy-flavor hadrons, some transport models are popular (see, e.g.,~reference~\cite{He:2014cla} and references therein).
In particular, the~coalescence or recombination models also provide good descriptions of hadron production especially at low and intermediate transverse momenta~\cite{Greco:2003xt,Fries:2003vb,Molnar:2003ff,Hwa:2002tu,Shao:2004cn,Greco:2003vf,Oh:2009zj,Liu:2012tn}.


Based on Qu-Bing Xie's works in $e^{+}e^{-}$ and pp collisions in early years~\cite{Xie:1984uoa,Xie:1988wi,Xie:1997ap,Chen:1988qi,Fang:1989ex,Liang:1991ya}, we developed a quark combination model (QCM) for hadronization and it works well in explaining yields, rapidity distributions, and transverse momentum spectra for the identified hadrons in high-energy heavy ion collisions at various energies ranging from RHIC to LHC~\cite{Shao:2004cn,Shao:2009uk,Wang:2013pnn,Song:2013isa}. Recently, inspired by the property of constituent quark number scaling for transverse momentum spectra of strange hadrons in p+Pb collisions at LHC energy~\cite{Song:2017gcz}, we proposed a simplified version of the quark combination model by incorporating the equal-velocity combination (EVC) to replace the near rapidity combination in the original model. Many properties of hadron production can be analytically derived and some of them have been tested by experimental data in high energy pp, pA, and AA collisions~\cite{Gou:2017foe,Wang:2019fcg,Song:2019sez,Li:2021nhq,Song:2021ojn}. Furthermore, our studies show that the EVC of charm and light quarks can explain the transverse momentum spectra of single-charm hadrons at low and intermediate transverse momenta~\cite{Song:2018tpv,Li:2017zuj,Wang:2019fcg,Li:2021nhq,Song:2021ojn}. In~particular, the~model prediction of the $\Lambda_c^+/D^0$ ratio was verified by the latest measurements of the ALICE collaboration~\cite{ALICE:2021rzj,ALICE:2022ych,ALICE:2018hbc}.


Recently, the ALICE collaboration published precise measurements of strange and charm hadrons, especially $D$ mesons and $\Lambda_c^+$ baryons, in~Pb+Pb collisions at LHC~\cite{ALICE:2018lyv,ALICE:2020try,ALICE:2021rxa,ALICE:2021kfc,ALICE:2021bib}.
In this paper, we apply the QCM with EVC to study the production of strange and charm hadrons simultaneously at low and intermediate transverse momenta in Pb+Pb collisions at $\sqrt{s_{NN}}=$5.02 TeV. We will present analytical and numerical results for the $p_T$ dependence of production ratios between different strange and charm hadrons. We will compare our results with the experimental data available and make predictions for other types of~hadrons.


The rest of the paper is organized as follows.
In Section~\ref{model}, we introduce a general phase-space structure of the QCM in heavy ion collisions as well as the idea and formula of the QCM in momentum space based on EVC. In~Sections~\ref{pt-strange} and \ref{pt-charm}, we apply the QCM to calculate spectra of various strange and charm hadrons in Pb+Pb collisions at \linebreak $\sqrt{s_{NN}}=$5.02 TeV and compare them with data.  The~final section is a summary of the main results and~conclusions.

\section{the Quark Combination~Model}
\label{model}

The QCM developed by the Shandong group led by Qu-Bing Xie~\cite{Xie:1984uoa,Xie:1988wi,Xie:1997ap,Chen:1988qi,Fang:1989ex,Liang:1991ya} is a kind of exclusive or statistical hadronization model with constituent quarks as building blocks. A~quark combination rule (QCR) can be derived for quarks and antiquarks in the neighborhood of the longitudinal phase space (momentum rapidity) to combine into baryons and mesons~\cite{Xie:1984uoa,Xie:1988wi,Xie:1997ap,Chen:1988qi,Yang:2019bbb}.
The QCM based on the QCR has successfully explained experimental data on hadron production in $e^{+}e^{-}$ and pp collisions~\cite{Xie:1984uoa,Xie:1988wi,Xie:1997ap,Chen:1988qi,Fang:1989ex,Liang:1991ya,Wang:1995ch} as well as in heavy ion collisions~\cite{Shao:2009uk,Shao:2004cn,Wang:2012cw}.
A modern version of QCM with spin degrees of freedom in terms of Wigner functions has been developed by some of us and applied to spin polarization of hadrons in heavy ion collisions~\cite{Liang:2004xn,Yang:2017sdk,Sheng:2019kmk,Sheng:2022wsy}.

In this section, we introduce the general phase-space structure of the QCM in heavy ion collisions as well as its simplified version in momentum space to describe momentum spectra of strange and charm~hadrons.


\subsection{General Phase Space Structure of QCM in Heavy Ion~Collisions}
\label{pssqcm}

In the quantum kinetic theory, the~formation of a composite particle through
the coalescence or combination process of its constituent particles $q_{1}q_{2}\cdots q_{n}\rightarrow H$
can be described by the collision term incorporating the matrix element
squared of the process and momentum integrals. In~the case we are considering, the~composite
particle $H$ can be a meson or a baryon, so the constituent particles $q_{1}q_{2}\cdots q_{n}$ are a quark
and an antiquark for the meson, and~are three quarks or three antiquarks
for the baryon or antibaryon, respectively. In~heavy ion collisions,
the coalescence process takes place in a space-time region, i.e.,~the freeze-out hypersurface
defined by the proper time $\tau_{0}$. The~momentum distribution of the hadron
(meson or baryon) reads
\begin{eqnarray}
    &&{}f_{H}(\mathbf{p}) \nonumber \\
    &\sim & \int d\sigma^{\mu}p_{\mu}\int\prod_{i=1}^{n}\frac{d^{3}\mathbf{p}_{i}}{(2\pi)^{3}2E_{i}} (2\pi)^{4}\delta(p_{1}+p_{2}+\cdots+p_{n}-p) \nonumber \\
    &\times&\left|M(q_{1}q_{2}\cdots q_{n}\rightarrow H)\right|^{2}f_{1}(x,p_{1})f_{2}(x,p_{2})\cdots f_{n}(x,p_{n})  
\label{general-form}
\end{eqnarray}
where $p=(E_{p},\mathbf{p})$ is the hadron's on-shell momentum, $p_{i}=(E_{i},\mathbf{p}_{i})$
is the on-shell momentum of the constituent particle $q_{i}$ with
its momentum distribution $f_{i}(x,p_{i})$ at the space-time point
$x$ on the freeze-out hypersurface, $M$ is the invariant amplitude of the coalescence process
containing the hadron's wave function, and~$d\sigma^{\mu}(x)$ is
the surface element pointing to the normal direction of the freeze-out
hypersurface at $x$. The~momentum distribution can be decomposed
into the thermal part and non-thermal part,
\begin{equation}
f_{i}(x,p_{i})=f_{i}^{\mathrm{th}}(\beta u\cdot p_{i})+f_{i}^{\mathrm{nth}}(p_{i}),
\end{equation}
where the thermal part $f_{i}^{\mathrm{th}}$ depends on $\beta u\cdot p_{i}$
with $\beta(x)=1/T(x)$ being the inverse temperature and $u^{\mu}(x)$
being the flow velocity both of which are functions of $x$ on the
freeze-out hypersurface, and~the non-thermal part $f_{i}^{\mathrm{nth}}$ depends
only on momentum and is independent of the space-time coordinate.
We can express the space-time point on the freeze-out hypersurface
in terms of the proper time $\tau$ and space-time rapidity $\eta$
as
\begin{equation}
x^{\mu}=(\tau\cosh\eta,\mathbf{x}_{T},\tau\sinh\eta),
\end{equation}
and also the hadron's on-shell momentum in terms of transverse momentum
$\mathbf{p}_{T}$ and rapidity $Y$ as
\begin{equation}
p^{\mu}=\left(m_{T}\cosh Y,\mathbf{p}_{T},m_{T}\sinh Y\right),
\end{equation}
where $m_{T}=\sqrt{m^{2}+p_{T}^{2}}$ is the transverse mass. Then
the freeze-out hypersurface element can be expressed as
\begin{equation}
d\sigma^{\mu}=\tau d\eta d^{2}x_{T}\frac{\partial x^{\mu}}{\partial\tau}=\tau d\eta d^{2}x_{T}(\cosh\eta,0,0,\sinh\eta),
\end{equation}
so its contraction with the hadron's momentum reads
\begin{equation}
d\sigma^{\mu}p_{\mu}=\tau d\eta d^{2}x_{T}m_{T}\cosh(\eta-Y).
\end{equation}

We can express the flow velocity with Bjorken's boost invariance
in the longitudinal direction with $\eta=\eta_{\mathrm{flow}}$,
\begin{align}
u^{\mu}(x)= & \left[\cosh\eta\cosh\rho(x_{T},\phi_{s}),\sinh\rho(x_{T},\phi_{s})\cos\phi_{b},\right.\nonumber \\
 & \left.\sinh\rho(x_{T},\phi_{s})\sin\phi_{b},\sinh\eta\cosh\rho(x_{T},\phi_{s})\right],
\end{align}
where $\rho(x_{T},\phi_{s})$ is the transverse flow rapidity~\cite{Wiedemann:1995au,Retiere:2003kf} as a
function of cylindrical coordinates in the transverse plane $x_{T}=|\mathbf{x}_{T}|$
and $\phi_{s}$, and~$\phi_{b}$ is the boost angle in the transverse
plane which can simply be taken as $\phi_{s}$ in approximation. The~elliptic flow can be implemented by~\cite{Retiere:2003kf}
\begin{equation}
\rho(x_{T},\phi_{s})=\frac{x_{T}}{R}\left[\rho_{0}+\rho_{2}\cos(2\phi_{s})\right],
\end{equation}
where $R$ is the transverse size of the fireball, and~$\rho_{2}$
is linked to the elliptic flow coefficient $v_{2}$.


\subsection{QCM in Momentum Space with Equal-Velocity~Combination}

For the purpose of this paper, we will introduce a simplified version of the QCM
in momentum space with an equal-velocity combination for hadron production.
This corresponds to (a) the quark distributions are homogeneous in space-time and depend only on momentum
and (b) the role of the matrix element squared is taken by the EVC. This version of QCM
is an approximation to the rigorous one in Subsection \ref{pssqcm}.

We consider a color-neutral system of $N_q=\sum _i N_{q_i}$ quarks
and $N_{\bar{q}}=\sum _i N_{\bar{q}_i}$ antiquarks
where $q_i=u,d,s,c$ and $\bar{q}_i=\bar{u},\bar{d},\bar{s},\bar{c}$ denote the quark and
antiquark flavors, respectively. The~momentum distributions $f_{M_j}(p)\equiv f_{M_j}(p;N_{q},N_{\bar{q}})$ and $f_{B_j}(p) \equiv f_{B_j}(p;N_{q},N_{\bar{q}})$
for the directly produced meson $M_j$ and baryon $B_j$ by combining a pair of quark--antiquarks and three quarks, respectively, can be schematically expressed as
\begin{eqnarray}
    &&{}f_{M_j}(p)\nonumber \\
    &=&\sum\limits_{\bar{q}_{1}q_{2}}  \int dp_1 dp_2 N_{\bar{q}_{1}q_{2}} f^{(n)}_{\bar{q}_{1}q_{2}}(p_1,p_2)
\mathcal {R}_{M_j,\bar{q}_{1}q_{2}}(p;p_1,p_2), \ \    \label{eq:fMjgeneral}     \\
    &&{} f_{B_j}(p) \nonumber \\
    &=&\sum\limits_{q_{1}q_{2}q_{3}}  \int dp_1 dp_2 dp_3
N_{q_1q_2q_3} f^{(n)}_{q_{1}q_{2}q_{3}}(p_1,p_2,p_3) \nonumber \\
    &\times& \mathcal {R}_{B_j,q_{1}q_{2}q_{3}}(p;p_1,p_2,p_3),
\label{eq:fBjgeneral}
\end{eqnarray}
where $f^{(n)}_{\bar{q}_{1}q_{2}}$ and $f^{(n)}_{q_{1}q_{2}q_{3}}$ are normalized joint momentum distributions;
$N_{\bar{q}_{1}q_{2}}$ and $N_{q_1 q_2 q_3}$ are the number of $\bar q_1 q_2$ pairs
and that of $q_1 q_2 q_3$ clusters in the system;
$\mathcal {R}_{M_j,\bar{q}_{1}q_{2}}$ and $\mathcal {R}_{B_j,q_{1}q_{2}q_{3}}$
are combination kernel functions that stand for the probability density
for a $\bar q_1q_2$ pair with momenta $p_1$ and $p_2$
to combine into a meson $M_j$ of momentum $p$ and that for
a $q_1q_2q_3$ cluster with $p_1$, $p_2$, and $p_3$ to combine into a baryon $B_j$ of momentum $p$, respectively.

Just as derived in Refs.~\cite{Song:2019sez,Wang:2019fcg}, the~combination kernel functions in the EVC can be written as
{\setlength\arraycolsep{0.2pt}
\begin{eqnarray}
    &&{}\mathcal{R}_{M_j,\bar q_1q_2} (p;p_{1},p_{2}) \nonumber \\
&=& C_{M_j}  \mathcal {R}_{\bar q_{1} q_2}^{(f)} \mathcal A_{M,{\bar q_1q_2}}
\delta(p_1-x_{q_1q_2}^{q_1}p) \delta(p_2-x_{q_1q_2}^{q_2}p),  \ \  \label{eq:RMj}  \\
    &&{}\mathcal {R}_{B_j,q_1q_2q_3} (p;p_{1},p_{2},p_{3}) \nonumber \\
&=& C_{B_j}   \mathcal {R}_{q_{1}q_{2}q_3}^{(f)} \mathcal A_{B,{q_1q_2q_3}}
\delta(p_1-x_{q_1q_2q_3}^{q_1}p)   \nonumber  \\
&&\times \delta(p_2-x_{q_1q_2q_3}^{q_2}p) \delta(p_3-x_{q_1q_2q_3}^{q_3}p)\,,
\label{eq:RBj}
\end{eqnarray} }%
where $\delta$-functions guarantee the momentum conservation in the EVC and~$x^{q_i}_{q_1q_2}=m_{q_i}/(m_{q_1}+m_{q_2})$ and
$x^{q_i}_{q_1q_2q_3}=m_{q_i}/(m_{q_1}+m_{q_2}+m_{q_3})$ are the momentum fraction of
the produced hadron for $q_i$. We note that the mass fraction is the same as the momentum fraction in the EVC. {Masses of up, down, strange, and charm quarks are taken to be $m_u=m_d=0.3$ GeV, $m_s=0.5$ GeV and $m_c=1.5$ GeV, respectively. }


The factor $C_{M_j}$ is the probability for $M$ to be $M_j$
if the quark content of $M$ is the same as $M_j$ and similar for $C_{B_j}$.
In this paper, we only consider hadrons in the ground state, namely mesons with $J^P=0^-$ and $1^-$
and baryons with $J^P=(1/2)^+$ and $(3/2)^+$. In~this case, $C_{M_j}$ is the same for all hadrons
in the same multiplet (with the same $J^P$) and determined by the production ratio of vector to pseudo-scalar mesons $R_{V/P}$, so is it for $C_{B_j}$ which is determined by the production ratio of
$J^P=({1}/{2})^+$ to $J^P=({3}/{2})^+$ baryons $R_{O/D}$ with the same flavor~content.


The factors $\mathcal {R}_{\bar q_1q_2}^{(f)}$ and $\mathcal {R}_{q_{1}q_{2}q_3}^{(f)}$
contain Kronecker $\delta$'s to guarantee the quark flavor conservation, e.g.,~if $M_j$ is a $D$-meson with constituent quark content $\bar q c$, $\mathcal {R}^{(f)}_{\bar q_{1} q_2}=\delta_{q_1,q} \delta_{q_2,c}$.
If $B_j$ is a single-charm baryon with the quark content $u d c$,
$\mathcal {R}_{q_{1}q_{2}q_3}^{(f)}=N_{\mathrm{sym}}\delta_{q_1,u} \delta_{q_2,d} \delta_{q_3,c}$,
where $N_{\mathrm{sym}}=1,3,6$ is a symmetry factor to account for
the number of different permutations of three quarks for (a) three identical flavors,
(b) two identical flavors, and~(c) all three distinct flavors, respectively.


The factor $\mathcal A_{M,{\bar q_1q_2}}$ is the probability
for a quark $q_2$ to capture a specific antiquark $\bar q_1$
to form a meson in the quark--antiquark system; it should be inversely proportional to $N_q+N_{\bar q}$.
Similarly, $\mathcal A_{B,{q_{1}q_{2}q_{3}}}$ should be inversely proportional to ${(N_q+N_{\bar q})^2}$.
Both $\mathcal A_{M,{\bar q_1q_2}}$ and $\mathcal A_{B,{q_{1}q_{2}q_{3}}}$ are
determined by the unitarity and the competition mechanism of meson-baryon production.
Note that for light-quark systems produced in $e^+e^-$ and pp collisions,
$\mathcal A_{M,{\bar q_1q_2}}$ and $\mathcal A_{B,{q_{1}q_{2}q_{3}}}$ correspond
to combination weights of mesons and baryons that follow the QCR~\cite{Xie:1984uoa,Xie:1988wi,Yang:2019bbb}.


Putting all these factors together, for~charm hadrons we are considering, \linebreak \mbox{Equations~(\ref{eq:fMjgeneral}) and (\ref{eq:fBjgeneral})} become
\begin{eqnarray}
f_{M_j}(p) &=& \frac{N_{c}N_{\bar{q}_1}}{N_q+N_{\bar{q}}} \mathcal{A}_{M} C_{M_j}
f^{(n)}_{\bar q_1 c} (x_{q_1c}^{q_1} p,x_{q_1c}^{c} p),
\label{eq:fMj}     \\
f_{B_j}(p) &=& \frac{N_{c}N_{q_1}N_{q_2}}{(N_q+N_{\bar{q}})^2}
\mathcal{A}_{B} C_{B_j} N_{\mathrm{sym}} \nonumber \\
    &\times& f^{(n)}_{q_1q_2c}(x_{q_1q_2c}^{q_1}p,x_{q_1q_2c}^{q_2}p,x_{q_1q_2c}^{c}p)\,,
\label{eq:fBj}
\end{eqnarray}
where $\mathcal{A}_{M}$ and $\mathcal{A}_{B}$ are two global coefficients
that can be determined by quark number conservation in the combination process
and the baryon-to-meson production ratio $N_{B}/N_M$.
We are considering a quark--antiquark system in the mid-rapidity region at very high collision energies,
so that net baryon number and net quark flavor are negligible, \linebreak i.e.,~$N_{\bar q_i} \approx N_{q_i}$
for $i=u,d,s,c$. Moreover, we assume that the number of strange quarks is suppressed
by a factor $\lambda_s$ (strangeness suppression factor) relative to that of up and down quarks,
so we have $N_u:N_d:N_s=1:1:\lambda _s$.


If we neglect correlations in the joint momentum distributions among different momenta,
we have factorization forms for the joint momentum distributions,
\begin{eqnarray}
f^{(n)}_{\bar{q}_{1}q_{2}}(p_1,p_2)
&=& f^{(n)}_{\bar{q}_{1}}(p_1) f^{(n)}_{q_{2}}(p_2),
\label{eq:fnq1q2}   \\
f^{(n)}_{q_{1}q_{2}q_{3}}(p_1,p_2,p_3)
&=&f^{(n)}_{q_{1}}(p_1)  f^{(n)}_{q_{2}}(p_2)  f^{(n)}_{q_{3}}(p_3).
\label{eq:fnq1q2q3}
\end{eqnarray}
We will use the above factorization forms in Equations~(\ref{eq:fMj},\ref{eq:fBj}) in our numerical calculation for single-charm hadrons. By~using Equations~(\ref{eq:fMj},\ref{eq:fBj}) with Equations~(\ref{eq:fnq1q2},\ref{eq:fnq1q2q3}), we are able to calculate momentum spectra and yields for different~hadrons.


Including strong and electromagnetic decay contributions from short-lived resonances~\cite{ParticleDataGroup:2018ovx},
we can obtain the momentum spectra of final state hadrons and make comparison with
experimental data. For~charm hadrons, we make an approximation that the momentum of the daughter charm hadron is
almost equal to that of the mother charm hadron.
With this approximation and the production ratio of the vector to the pseudo-scalar meson being set to 1.5~\cite{ALICE:2012inj,Li:2017zuj}, we obtain (for the final state $D$ mesons):
\begin{eqnarray}
f_{D^0}^{(\mathrm{fin})}(p) &\approx& 3.516 f_{D^0}(p),  \label{eq:D0fin}   \\
f_{D^+}^{(\mathrm{fin})}(p)  &\approx& 1.485 f_{D^+}(p),  \label{eq:Dpfin}  \\
f_{D_s^+}^{(\mathrm{fin})}(p)  &\approx& 2.5 f_{D_s^+}(p).  \label{eq:Dspfin}
\end{eqnarray}

Similarly, we can set the production ratio of $J^P=(1/2)^+$ to the $J^P=(3/2)^+$ single-charm baryon to 2~\cite{Li:2017zuj}
and obtain,
\begin{eqnarray}
f_{\Lambda_c^+}^{(\mathrm{fin})}(p)  &\approx& 5 f_{\Lambda_c^+}(p),  \label{eq:Lamcpfin} \\
f_{\Sigma_c^0}^{(\mathrm{fin})}(p)  &\approx&  f_{\Sigma_c^0}(p),  \label{eq:Sigc0fin} \\
f_{\Sigma_c^+}^{(\mathrm{fin})}(p)  &\approx& f_{\Sigma_c^+}(p),  \label{eq:Sigcpfin} \\
f_{\Sigma_c^{++}}^{(\mathrm{fin})}(p)  &\approx& f_{\Sigma_c^{++}}(p),  \label{eq:Sigcppfin} \\
f_{\Xi_c^0}^{(\mathrm{fin})}(p)  &\approx& 2.5 f_{\Xi_c^0}(p),  \label{eq:Xic0fin} \\
f_{\Xi_c^+}^{(\mathrm{fin})}(p)  &\approx& 2.5 f_{\Xi_c^+}(p),  \label{eq:Xicpfin} \\
f_{\Omega_c^0}^{(\mathrm{fin})}(p)  &\approx& 1.5 f_{\Omega_c^0}(p).  \label{eq:Omec0fin}
\end{eqnarray}

These analytical results can be used to obtain the $p_T$ spectra of charm hadrons.
For final state strange hadrons, there are no such analytical results, only numerical~ones.




\section{Transverse Momentum Spectra and Baryon-To-Meson Ratio for Strange~Hadrons}
\label{pt-strange}

In this section, we apply the QCM introduced in Section~\ref{model}
to study the production of strange hadrons in Pb+Pb collisions at $\sqrt{s_{NN}}= 5.02$ TeV.
We first calculate the $p_T$ spectra of strange mesons and baryons.
Then we calculate the baryon-to-meson ratio $\Lambda/K_s^0$ as a function of $p_T$ in different~types of centralities.


\subsection{Transverse Momentum Spectra of Strange~Hadrons}

\begin{figure*}[!htpb]
 \includegraphics[width=0.7\linewidth]{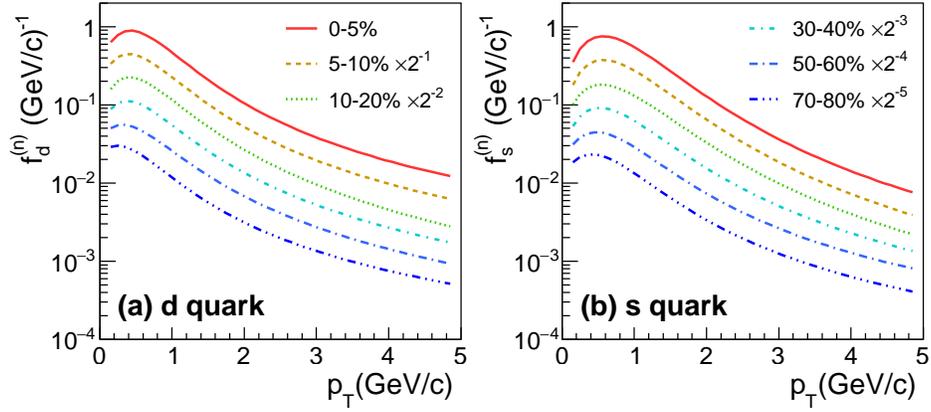}\\
 \caption{The normalized $p_T$ distributions of (a) $d$ quark and (b) $s$ quark in different centralities in Pb+Pb collisions at $\sqrt{s_{NN}}=5.02$ TeV.
}
\label{fig:dsquarkpt}
\end{figure*}

The inputs of the model are $p_T$ spectra of quarks and antiquarks.
In this paper, we adopt the isospin symmetry and neglect the net quark numbers ($N_{q_i}\approx N_{\bar{q}_i}$ for $i=u,d,s,c$) in the mid-rapidity region at LHC energy, so we have only two inputs $f_d(p_T)=f_u(p_T)$ and $f_s(p_T)$,
which can be fixed by fitting the experimental data on the $p_T$ spectra of $\phi$ mesons and $\Lambda$ baryons~\cite{ALICE:2019xyr,Kalinak:2017xll,ALICE:2021ptz}.
The extracted results for the normalized $p_T$ spectra of quarks in central 0-5\% to peripheral 70-80\% Pb+Pb collisions
at $\sqrt{s_{NN}}=$ 5.02 TeV are shown in Figure~\ref{fig:dsquarkpt}.
The rapidity densities of $d$ and $s$ quarks are listed in Table~\ref{tab_dsinputs}.

\begin{table}[H]
\renewcommand{\arraystretch}{1.5}
\centering
\caption{Rapidity densities of $d$ and $s$ quarks in different centralities in Pb+Pb collisions at \linebreak $\sqrt{s_{NN}}=$ 5.02~TeV.}
    \begin{tabular}{@{}lll@{}}
     \toprule
   \textbf{Centrality}~~~~~~~~~~~~~~      & \boldmath{$dN_d/dy$}~~~~~~~~~~~~~~~           &\boldmath{ $dN_s/dy$}                    \\   \midrule
   0-5\%        &840   &370                 \\
  5-10\%        &686   &302                 \\
  10-20\%        &516   &227                 \\
  30-40\%        &267   &115                 \\
  50-60\%        &97   &40                 \\
  70-80\%        &27   &10                 \\
  \bottomrule
\end{tabular}     \label{tab_dsinputs}
\end{table}

\begin{figure*}[!htpb]
\includegraphics[width=0.80\linewidth]{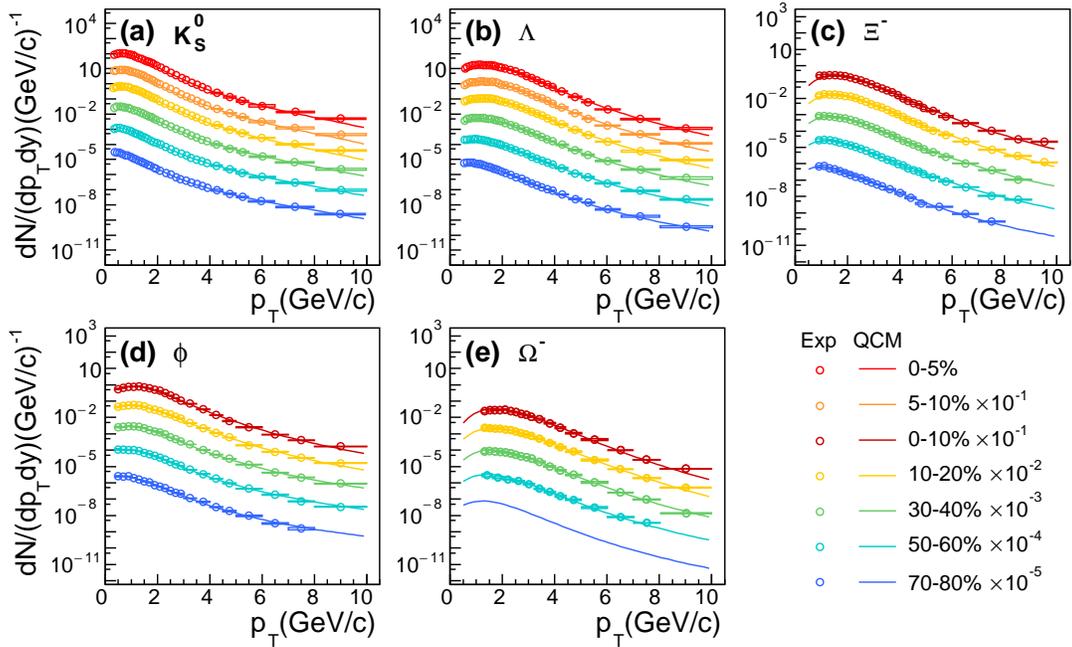}\\
\caption{The $p_T$ spectra of (a) $K_s^0$, (b) $\Lambda$, (c) $\Xi^-$, (d) $\phi$ and (e) $\Omega^-$ in different centralities in Pb+Pb collisions at $\sqrt{s_{NN}}=$5.02 TeV. Open symbols are experimental data~\cite{ALICE:2019xyr,Kalinak:2017xll}. { Lines for $\Lambda$ and $\phi$ are fitting results that are used to fix $p_T$ spectra of light-flavor quarks at hadronization in QCM. Lines for $K_{S}^{0}$, $\Xi^{-}$ and $\Omega^{-}$ are predictions from QCM. }}
\label{fig:sHpt}
\end{figure*}

In Figure~\ref{fig:sHpt}, we show the results for the $p_T$ spectra of $K_s^0$ and $\Lambda$ in 0--5\%, 5--10\%, 10--20\%, 30--40\%, 50--60\%, 70--80\% centralities, and those of $\phi$, $\Xi^-$ and $\Omega^-$ in 0--10\%, 10--20\%, 30--40\%, 50--60\%, 70--80\% centralities. The~QCM results are displayed in lines and experimental
data~\cite{ALICE:2019xyr,Kalinak:2017xll} are displayed in open symbols.
From Figure~\ref{fig:sHpt}, we see that our QCM results for strange mesons and baryons agree
with the experimental data very well.
Such a good agreement provides a piece of evidence for the EVC mechanism in describing strange hadron production
in Pb+Pb collisions at the LHC~energy.

\subsection{Baryon-To-Meson Ratio $ \Lambda/K_S^0$}

\begin{figure*}[!htpb]
\includegraphics[width=0.8\linewidth]{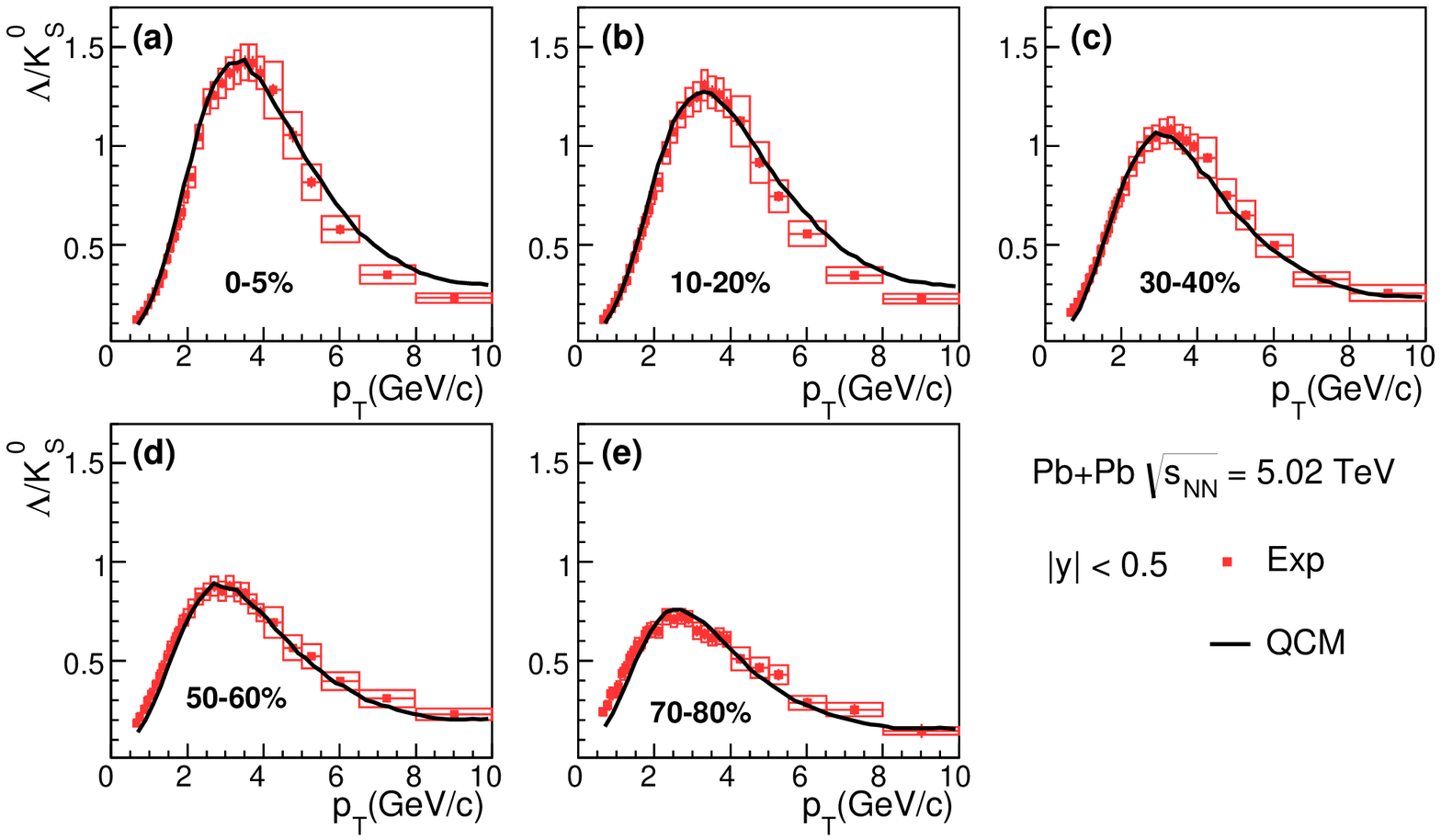}\\
\caption{The 
 $p_T$ dependence of $\Lambda/K_S^0$ in (\textbf{a}) 0--5\%, (\textbf{b}) 10--20\%, (\textbf{c}) 30--40\%, (\textbf{d}) 50-6-0\%, (\textbf{e}) 70--80\% centralities in Pb+Pb collisions at $\sqrt{s_{NN}}=$5.02 TeV.
Filled squares are experimental data 
~\cite{Kalinak:2017xll} and lines are the QCM results.}
\label{fig:RLamKs0}
\end{figure*}

Figure~\ref{fig:RLamKs0} shows the multiplicity ratio $\Lambda/K_S^0$ as a function of $p_T$ in five centrality ranges 0--5\%, 10--20\%, 30--40\%, 50--60\%, and~70--80\%. Filled squares are experimental data~\cite{Kalinak:2017xll},
and lines are the QCM results. We see that $\Lambda/K_S^0$ exhibits an increase-peak-decrease behavior
as a function of $p_T$ in all centralities, which is regarded as a natural consequence of
quark recombination~\cite{Hwa:2002tu,Greco:2003xt,Fries:2003vb,Greco:2003mm,Fries:2003kq}.
We see that this feature can be well described by the QCM with EVC.
The height of the peak increases from about 0.8 in the peripheral (70--80\% centrality) to about 1.5 in the most central
(0-5\% centrality) collisions. The~peak positions in the $p_T$ slightly move to higher values
from peripheral to central collisions due to stronger radial flows in more central collisions.
The QCM with EVC gives a good description of the $p_T$ dependence of $\Lambda/K_S^0$.

\section{Transverse Momentum Spectra, Yield Ratios, and Nuclear Modification Factor for Charm~Hadrons}
\label{pt-charm}

In this section, we apply the QCM with EVC to study the production of charm hadrons at midrapidity
in Pb+Pb collisions at $\sqrt{s_{NN}}=$5.02 TeV. We first calculate the $p_T$ spectra of $D$ mesons and single-charm baryons.
Then we give the $p_T$ dependence of yield ratios of different charm hadrons.
Finally, we give the nuclear modification factor $R_{AA}$ for charm hadrons as a function of $p_T$.


\subsection{Transverse Momentum Spectra of Charm Mesons and~Baryons}

In the QCM, the~only additional input is the normalized $p_T$ distribution of the charm quarks,
which we adopt a hybrid form based on the simulation of charm quarks propagating
in the QGP medium in a Boltzmann transport approach~\cite{Scardina:2017ipo,Fries:2003kq}
\begin{widetext}
\begin{equation}
    f^{(n)}_{c}(p_T)=\frac{1}{N_{\mathrm{norm}}}p_{T} \left[ \left(\frac{p_T}{p_{T0}}\right)^{\alpha_c}\exp\left( -\frac{\sqrt{p_T^2+m_{c}^2}}{T_c}\right) + \left(1.0+\frac{\sqrt{p_T^2+m_{c}^2}-m_c}{\Gamma_c} \right) ^{-\beta_c}\right]. \label{c_fpt}
\end{equation}
\end{widetext}

\begin{figure}[H]
 \includegraphics[width=0.9\linewidth]{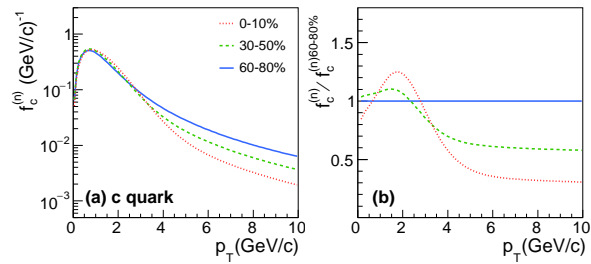}\\
 \caption{({\textbf{a})} 
 Normalized $p_T$ distribution of the charm quarks $f^{(n)}_{c}(p_T)$ at different centralities.
(\textbf{b}) The distribution $f^{(n)}_{c}(p_T)$ at different centralities normalized by that at 60--80\% centrality. }
 \label{fig:cquarkpt}
\end{figure}

{At small $p_T$, this parameterized form is very close to the thermal distribution, while at large $p_T$, it follows the power law which is a non-thermal distribution. 
Both the thermal and non-thermal distributions are smoothly connected through the above parameterization. }
Here, the~normalization constant $N_{\mathrm{norm}}$ can be determined by the condition $\int _0^{\infty} dp_T f^{(n)}_{c}(p_T)=1$.  The~parameters $\alpha_c$, $p_{T0}$, $T_c$, $\Gamma_c$, and $\beta_c$
are fitted using the data of $D^0$'s $p_T$ spectra~\cite{ALICE:2018lyv,ALICE:2021rxa}
and are listed in Table~\ref{tab_Cinputs}.
The shapes of $f^{(n)}_{c}(p_T)$ at different centralities are shown in Figure~\ref{fig:cquarkpt}a.
We see that there is a stronger suppression in more central collisions in the $p_T$ range 4 GeV $<p_T<$ 10 GeV.
Figure~\ref{fig:cquarkpt}b shows $f^{(n)}_{c}(p_T)$ at different centralities normalized by 60--80\% centrality,
which has similar behavior to the nuclear modification factor $R_{CP}$ of the $D^0$ meson measured in reference~\cite{ALICE:2018lyv}.
For the rapidity density of charm quarks ${dN_{c}}/{dy}$, we assume that it is proportional
to the cross-section per rapidity in pp collisions as
\begin{equation}
\frac{dN_{c}}{dy}=\langle T_{AA}\rangle \frac{d\sigma^{\mathrm{pp}}_{c}}{dy}.
\end{equation}
Here, $\langle T_{AA}\rangle$ is the average nuclear overlap function~\cite{ALICE:2018lyv},
${d\sigma^{pp}_{c}}/{dy}$ is the $p_T$ integrated cross-section of charm quarks in pp collisions which is about 1.0 mb at $\sqrt{s}=$5.02 TeV~\cite{Li:2021nhq}.
The values of ${dN_{c}}/{dy}$ at different centralities are listed in Table~\ref{tab_Cinputs}.

\begin{table}[H]

\caption{The parameters in the normalized $p_T$ distribution of the charm quarks at different~centralities. }
    \begin{tabular}{@{}llll@{}} 
    \toprule
   \textbf{Centrality}~~~~~~~~              &\textbf{ 0--10\% }~~~~~~~~~~~                 & \textbf{30--50\%}~~~~~~~~~                  & \textbf{60--80\% }    \\   
 $p_{T0}$ (GeV/c)              &0.0051   &0.10        &0.63                  \\
 $\alpha_c$             &0.5                         &1.0                          &1.5                   \\
 $T_c$ (GeV)          &0.46                       &0.38                        &0.34                \\
 $\beta_c$                &3.10                      &3.00                        &2.95               \\
$\Gamma_c$ (GeV)  &0.6                      &0.6                          &0.7               \\
 $dN_c/dy$              &23.07                   &3.90                        &0.417               \\
 \bottomrule
\end{tabular}
\label{tab_Cinputs}
\end{table}

\begin{figure*}[!htpb]
\includegraphics[width=0.70\linewidth]{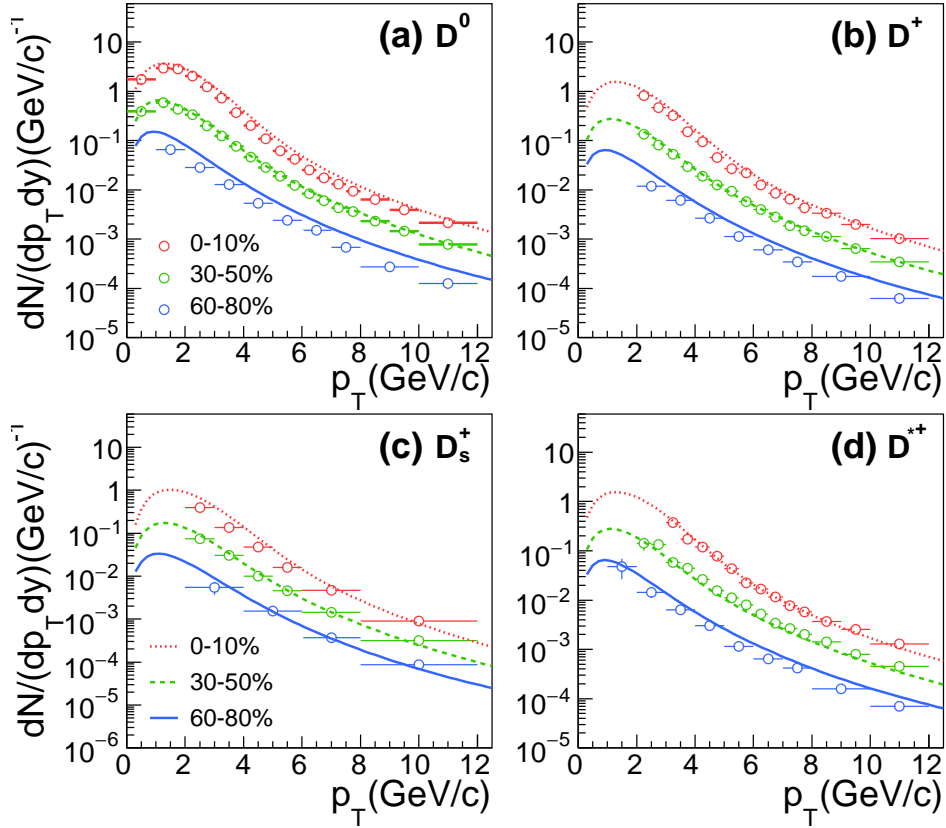}\\
    \caption{{The} 
 $p_T$ spectra of (\textbf{a}) $D^0$, (\textbf{b}) $D^+$, (\textbf{c}) $D^+_s$, and~(\textbf{d}) $D^{*+}$ mesons at different centralities.  Open symbols are {the experimental data} ~\cite{ALICE:2018lyv,ALICE:2021rxa,ALICE:2021kfc}. { Lines for $D^{0}$ are fitting results that are used to fix the shape parameters in the $p_T$ spectrum of charm quarks at hadronization in Equation~(\ref{c_fpt}). Lines for $D^{+}$, $D_{s}^{+}$ and $D^{*+}$ are predictions from QCM.} }
\label{fig:Dmesonpt}
\end{figure*}

In Figure~\ref{fig:Dmesonpt}, we present the results for the $p_T$ spectra of $D^0$, $D^+$, $D^+_s$, and~$D^{*+}$ mesons at 0--10\%, 30--50\%, and 60--80\% centralities. Open symbols are the {experimental data}
~\cite{ALICE:2018lyv,ALICE:2021rxa,ALICE:2021kfc} and lines are the calculated results. The~results agree with the experimental data in the $p_T$ range from 0.5 GeV/c up to 10 GeV/c.

\unskip

\begin{figure*}[!htpb]

\includegraphics[width=0.7\linewidth]{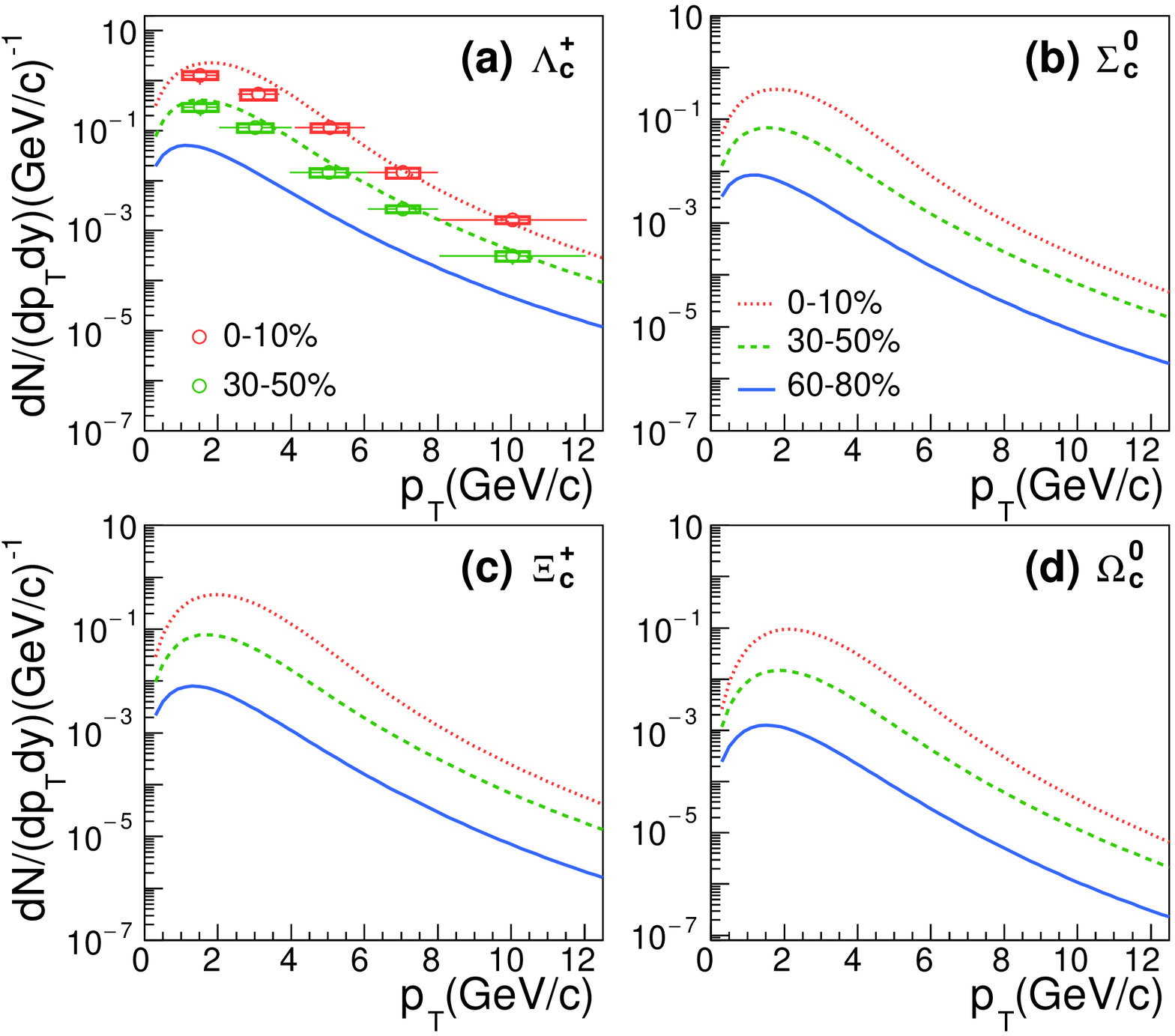}\\
\caption{{The} 
 $p_T$ spectra of (\textbf{a}) $\Lambda_c^+$, (\textbf{b}) $\Sigma_c^0$, (\textbf{c}) $\Xi_c^+$, and~(\textbf{d}) $\Omega_c^0$ baryons at different centralities. Open symbols are the data of $\Lambda_c^+$~\cite{ALICE:2021bib}, and~lines are the QCM~results. }
\label{fig:Bcpt}
\end{figure*}

Figure~\ref{fig:Bcpt} shows the results for charm baryons at 0--10\%, 30--50\%, and 60--80\% centralities.
Open symbols [only in Figure~\ref{fig:Bcpt}a] are the data of $\Lambda_c^+$~\cite{ALICE:2021bib} and lines are the results from the  QCM, which are in  good agreement. Predictions from QCM for other charm baryons $\Sigma_c^0$, $\Xi_c^+$, and~$\Omega_c^0$ are presented in Figure~\ref{fig:Bcpt}b--d, which can be tested in future experimental~measurements.

The agreement between our results and experimental data for various $D$ mesons and $\Lambda_c^+$ baryons indicates
the validity of the EVC mechanism in the QCM in describing the charm hadron production.
In this mechanism, the~formation of the charm hadron is through the capture of light quarks in the medium by the charm quark with the same~velocity.



We also calculated $p_T$-integrated yield density $dN/dy$ for charm hadrons
at midrapidity and 0--10\%, 30--50\%, and 60--80\% centralities as listed in Table~\ref{tab_yields}.
The experimental data are taken from Refs.~\cite{ALICE:2021rxa,ALICE:2021kfc}.
The QCM results are slightly higher than data.
This is because we only include single-charm hadrons in the calculation and assume all charm quarks go to single-charm hadrons.
In fact, the hidden-charm $J/\Psi$, double-charm baryons, and even heavy-flavor multiquark states can be produced.
If including these particles, our results in Table~\ref{tab_yields} will decrease slightly and the agreement with experimental data will be improved. No data are available for the yield densities of many charm hadrons; our QCM predictions can  be tested by their future experimental~measurements.


\begin{table*}[!htpb]
\centering
\caption{The yield density $dN/dy$ for charm hadrons at different centralities.
The experimental data are from Refs.~\cite{ALICE:2021rxa,ALICE:2021kfc}.\label{tab_yields} }
    \begin{tabular}{@{}p{30pt}p{140pt}p{25pt}p{5pt}p{140pt}p{25pt}p{5pt}p{25pt}p{30pt}@{}}
    \toprule
    \multirow{2}{*}{\textbf{Hadron}}   &\multicolumn{2}{c}{\textbf{0--10\%}}   &                                                     &\multicolumn{2}{c}{\textbf{30--50\%}}        &                            &\multicolumn{2}{c}{\textbf{60--80\%}}   \\     \cmidrule{2-3}     \cmidrule{5-6}   \cmidrule{8-9}
    &\textbf{Data}  &\textbf{QCM}  &                                                                             &\textbf{Data} &\textbf{QCM}  &                                                                        &\textbf{Data} & \textbf{QCM}   \\
    \midrule
    $D^0$        &$6.819\pm0.457^{+0.912}_{-0.936}\pm0.054$     &8.438&                                 &$1.275\pm0.099^{+0.167}_{-0.173}\pm0.010$  &1.436&                                     &---  &0.157       \\

    $D^+$       &$3.041\pm0.073^{+0.154}_{-0.155}\pm0.052^{+0.352}_{-0.618}$   &3.563&     &$0.552\pm0.008^{+0.024}_{-0.024}\pm0.009^{+0.068}_{-0.114}$  &0.606&      &---  &0.0665       \\

    $D^{*+}$  &$3.803\pm0.037^{+0.084}_{-0.085}\pm0.041^{+0.854}_{-1.175}$   &3.600&     &$0.663\pm0.023^{+0.038}_{-0.039}\pm0.007^{+0.149}_{-0.165}$ &0.613&       &--- &0.0672       \\

    $D_s^+$   &$1.89\pm0.07^{+0.13+0.36}_{-0.16-0.55}\pm0.07$     &2.417&                          &$0.34\pm0.01^{+0.02+0.11}_{-0.03-0.09}\pm0.01$    &0.395&                             &---     &0.0368       \\

    $\Lambda_c^+$     &--- &5.983&     &--- &1.026&        &--- &0.115        \\

    $\Sigma_c^0$        &--- &0.997&     &---&0.171&         &--- &0.0192        \\

    $\Sigma_c^{++}$  &--- &0.997&     &---&0.171&        &--- &0.0192      \\

    $\Xi_c^0$             &---&1.211&       &---  &0.199&      &---&0.0189      \\

    $\Xi_c^+$            &--- &1.211&       &---&0.199&        &--- &0.0189      \\

    $\Omega_c^0$     &--- &0.246&       &--- &0.0386&     &---  &0.00311     \\
    \bottomrule
\end{tabular}     
\end{table*}
\unskip


\subsection{Yield Ratios for Charm~Hadrons}

In this subsection, we calculate two kinds of yield ratios as functions of $p_T$ for charm hadrons:
one is $D_s^+/D^0$ which is related to strangeness production, and~the other is the baryon-to-meson~ratio.

\begin{figure*}[!htpb]
\includegraphics[width=0.92\linewidth]{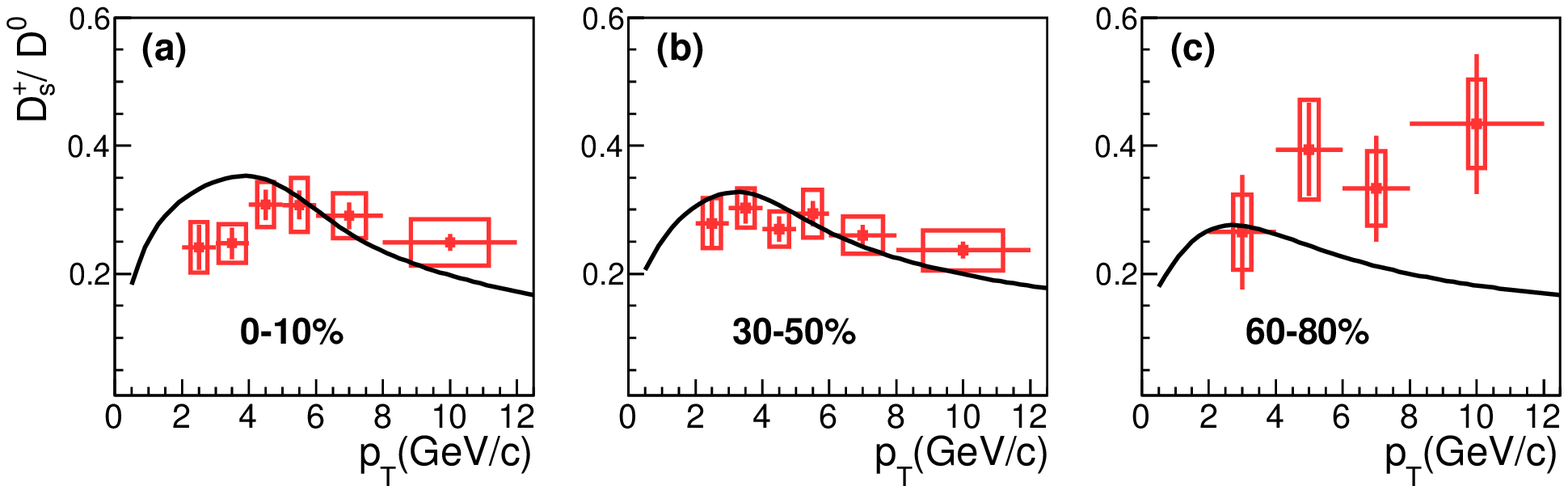}\\
\caption{$D_s^+/D^0$ as a function of $p_T$ for (\textbf{a}) 0--10\%, (\textbf{b}) 30--50\%, and (\textbf{c}) 60--80\% centralities.
The symbols are {experimental data}
~\cite{ALICE:2018lyv,ALICE:2021kfc}, and~solid lines are the QCM results. }
\label{fig:Rstrange-pt}
\end{figure*}

We first look at the results for $D_s^+/D^0$ in Figure~\ref{fig:Rstrange-pt}
at 0--10\%, 30--50\% and 60--80\% centralities. The~symbols in panels (a) and (b)
are from the most recent data~\cite{ALICE:2021kfc}, while those in panel (c) are from previous measurements~\cite{ALICE:2018lyv}.
Different lines are the QCM results.
The agreement between the data and our results with the same value of $\lambda_s$ extracted from strange hadrons
implies the same 'strangeness' environment for both strange and charm hadrons and supports
the QCM works as the hadronization mechanism for charm quarks in the QGP~medium.

We then look at the baryon-to-meson ratio $\Lambda_c^+/D^0$, which is considered a probe
to the charm quark hadronization. Recalling Equations~(\ref{eq:D0fin},\ref{eq:Lamcpfin}), we have
\begin{equation}
\frac{\Lambda_c^+}{D^0} = \frac{4.267}{2+\lambda_s}\cdot \frac{\mathcal A_{B}}{\mathcal A_{M}}
\cdot\frac{[f^{(n)}_{d}(x^{d}_{ddc}p_T)]^2  f^{(n)}_{c}(x^{c}_{ddc}p_T)}
{f^{(n)}_{d}(x^{d}_{dc}p_T) f^{(n)}_{c}(x^{c}_{dc}p_T)}\,.
\label{eq:RLcpD0fin}
\end{equation}

\begin{figure*}[!htpb]

\includegraphics[width=0.89\linewidth]{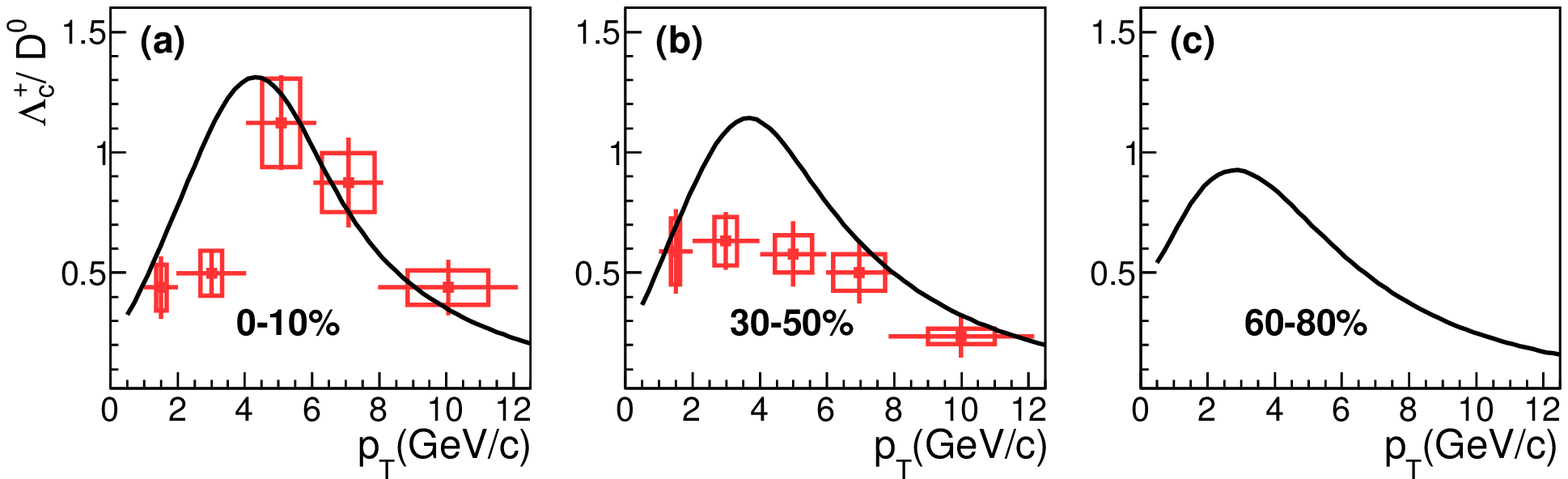}\\
\caption{The baryon-to-meson ratio $\Lambda_c^+/D^0$ as a function of $p_T$ for (\textbf{a}) 0--10\%, (\textbf{b}) 30--50\%, and \mbox{(\textbf{c}) 60--80\%} centralities. {Symbols are experimental data}
~\cite{ALICE:2021bib}, and~solid lines are the QCM~results. }
\label{fig:LcD0ratio-pt}
\end{figure*}

In our calculation, we set ${\mathcal A_{B}}/{\mathcal A_{M}}$ to 0.45 by $R_{B/M}=0.6$ in the charm sector~\cite{Wang:2019fcg}.
Following Equation~(\ref{eq:RLcpD0fin}), the~results for $\Lambda_c^+/D^0$ as a function of $p_T$ at different centralities
are given in Figure~\ref{fig:LcD0ratio-pt}. We see that $\Lambda_c^+/D^0$ as a function of $p_T$ shows a similar shape to $\Lambda/K_S^0$ in Figure~\ref{fig:RLamKs0} except a larger shift in the $p_T$ at the peak value from the central to peripheral collisions;
in the central to peripheral collisions, the~peak values decrease from about 1.3 to 0.9 and their locations in the $p_T$ shift from 5 to 3 GeV. The~peak locations shifting lower $p_T$ from the central to peripheral collisions is due to
the stronger collectivity in more central~collisions.

 
\unskip


\subsection{Nuclear Modification Factor $R_{AA}$}


We finally investigate the nuclear modification factor $R_{AA}$ for charm hadrons, {which is defined as }
\begin{equation}
    R_{AA}(p_T) = \frac{1}{\langle T_{AA}\rangle} \frac{dN_{AA}/dp_{T}}{d\sigma_{pp}/dp_T}.
\end{equation}

\begin{figure*}[!htpb]

\includegraphics[width=0.7\linewidth]{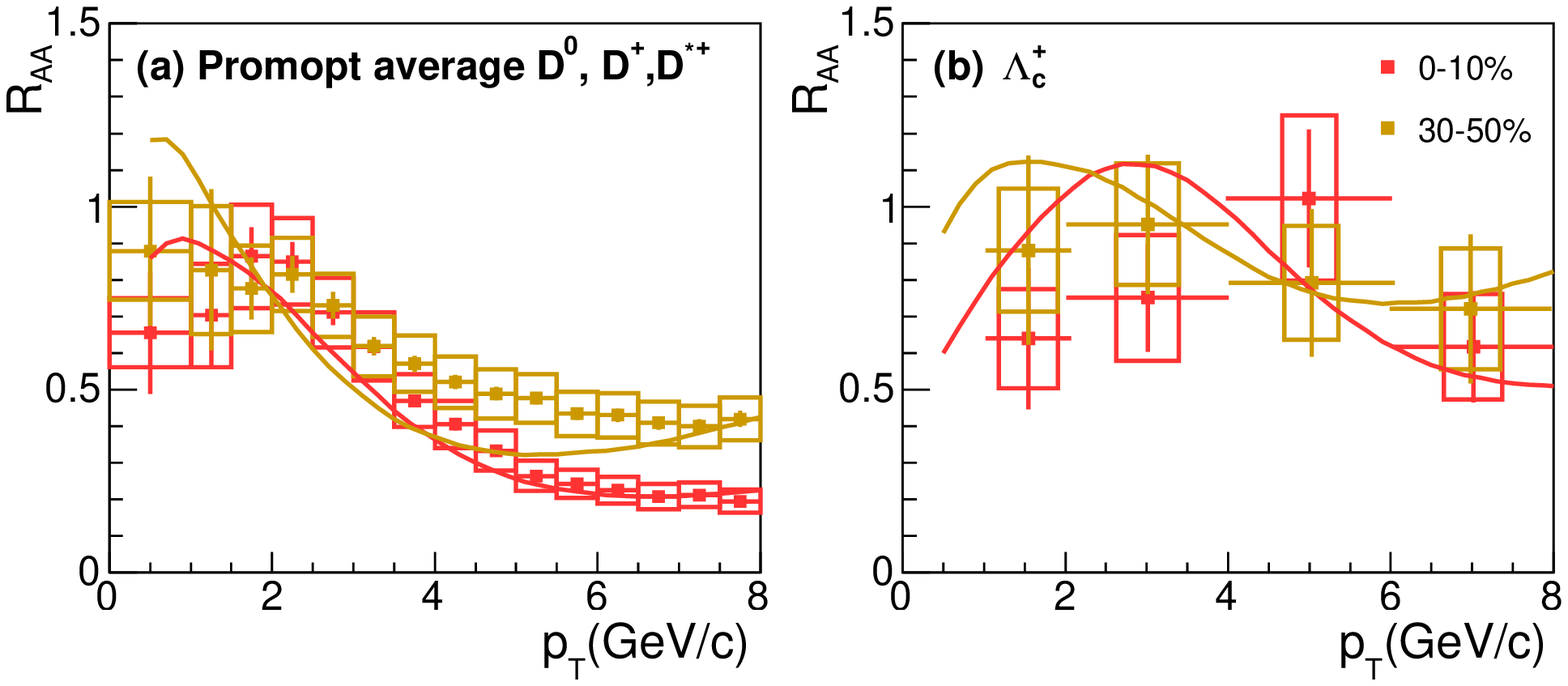}\\
\caption{The nuclear modification factor $R_{AA}$ for (\textbf{a}) prompt $D^0$, $D^+$ and $D^{*+}$ mesons and (\textbf{b}) $\Lambda_c^+$ baryons as functions of $p_T$ at different centralities. Symbols are {experimental data from Refs.}
~\cite{ALICE:2021rxa,ALICE:2021bib}, and~solid lines are the QCM~results. }
\label{fig:LcD0-RAA}
\end{figure*}

{The results for differential cross-sections of charm hadrons in $pp$ collisions at \linebreak $\sqrt{s}=5.02$ TeV are taken from reference~\cite{Li:2021nhq} by some of us. }
Figure~\ref{fig:LcD0-RAA} shows $R_{AA}$ for prompt $D^0$, $D^+$ and $D^{*+}$ mesons as well as
$\Lambda_c^+$ at 0--10\% and 30--50\% centralities.
Symbols are experimental data~\cite{ALICE:2021rxa,ALICE:2021bib}, and~solid lines are the QCM results.
One can see in Figure~\ref{fig:LcD0-RAA} that both $D$ mesons and $\Lambda_c^+$ have similar $p_T$ behaviors.
The peaks are located at $p_T^{\mathrm{peak}}$ which shifts towards higher values from peripheral to central collisions.
 {This shift is mainly due to the stronger collectivity in central collisions which can boost thermal quarks to larger transverse momenta that are passed to charm hadrons by the EVC mechanism.  We also see that the peak shift for $R_{AA}$ of $\Lambda_{c}^{+}$ is more obvious than that of $D$ mesons, because~$\Lambda_{c}^{+}$ contains two light quarks and therefore is more influenced by centrality-dependent collectivity. }

\section{Summary}

The comparative study of the production properties of strange and charm hadrons can provide information on the hadronization mechanism in relativistic heavy ion collisions.  We use a quark combination model in momentum space with the approximation of equal-velocity combination to study these properties in Pb+Pb collisions at $\sqrt{s_{NN}}=$5.02 TeV.
{We used experimental data of $\Lambda$, $\phi$ and $D^{0}$ to fix the $p_T$ spectra of up, strange, and charm quarks at hadronization. We computed the $p_T$ spectra and rapidity densities of $K_s^0$, $\Xi^-$, $\Omega^-$, $D^+$, $D^+_s$, $D^{*+}$, $\Lambda_c^+$, $\Sigma_c^0$, $\Xi_c^+$ and $\Omega_c^0$ from central to peripheral collisions.}
The QCM results agree with the available experimental data quite~well.


We calculated the yield ratios $\Lambda/K_S^0$, $D_s^+/D^0$ and $\Lambda_c^+/D^0$ as functions of $p_T$.
We found that the EVC-based QCM in momentum space can naturally describe their non-trivial behaviors as functions of the $p_T$ and the centrality.
The $p_T$ locations of the peaks in these ratio curves shift to lower values from central to peripheral collisions, an~effect arising from collectivity in heavy ion collisions, absent in pp collisions.
The calculated results for the nuclear modification factor $R_{AA}$ show similar behaviors for both $D$ and $\Lambda_c^+$, which can be tested by more precise experimental measurements, especially at a low $p_T$.
All our results support the validity of the EVC-based QCM in describing the hadronization mechanism of charm quarks in high energy heavy ion~collisions.


\section{Acknowledgments}
We dedicate this work to Qu-bing Xie (1935-2013) who was the teacher, mentor and friend of ZTL, FLS and QW.  This research was funded by the National Natural Science Foundation of China under grant nos. 12175115, 11975011, 12135011, 11890710 and 11890713, and~the Natural Science Foundation of Shandong Province, China, under~grant nos. ZR2019MA053 and ZR2020MA097.

\bibliographystyle{apsrev4-1}
\bibliography{ref}

\end{document}